\documentclass[usenatbib]{mnras}
\usepackage[T1]{fontenc}
\usepackage[latin9]{inputenc}
\usepackage{amsmath}
\usepackage{multirow}
\usepackage{graphicx}
\providecommand{\tabularnewline}{\\}
\begin{document}
\title{%Zonal harmonics 
Solar large-scale magnetic field and 
cycle patterns in solar dynamo}
\author[V.N.~Obridko, V.V.~Pipin, D.D.~Sokoloff, A.S.~Shibalova]
{V.N. Obridko$^{1}$, \thanks{email:obridko@izmiran.ru},
V.~Pipin$^{4}$\thanks{email: pip@iszf.irk.ru},
D.~Sokoloff$^{1,2,3}$\thanks{email: sokoloff.dd@gmail.com},
A.S. Shibalova$^{2,3}$,\thanks{email:as.shibalova@physics.msu.ru}\\
$^{1}${IZMIRAN, 4 Kaluzhskoe  Shosse, Troitsk, Moscow, 108840}\\
$^{2}${Department of Physics, Moscow State University,
  Moscow,119992, Russia}\\
$^{3}$Moscow Center of Fundamental and Applied Mathematics, Moscow,
119991, Russia\\
$^4${Institute of Solar-Terrestrial Physics, Russian Academy of
Sciences, Irkutsk, 664033, Russia} }
\maketitle 

\begin{abstract}
We compare spectra of the zonal harmonics of the large-scale magnetic
field of the Sun using observation results and solar dynamo models.
The main solar activity cycle as recorded in these tracers is a much
more complicated phenomenon than the eigen solution of solar dynamo
equations with the growth saturated by a back reaction of the dynamo-driven
magnetic field on solar hydrodynamics. The nominal 11(22)-year cycle
as recorded in each mode has a specific phase shift varying from cycle
to cycle; the actual length of the cycle varies from one cycle to
another and from tracer to tracer. Both the observation and the dynamo
model show an exceptional role of the axisymmetric $\ell_{5}$ mode.
Its origin seems to be readily connected with the formation and evolution
of sunspots on the solar surface. The results of observations and
dynamo models show a good agreement for the low $\ell_{1}$ and $\ell_{3}$
modes. The results for these modes do not differ significantly for
the axisymmetric and nonaxisymmetric models. Our findings support the idea
that the sources of the solar dynamo arise as a result of both the distributed dynamo processes in the bulk of the convection zone and the surface magnetic activity.
\end{abstract}

\section{Introduction}

Until the middle of the past century, the solar activity was considered
mainly as a periodic modulation of sunspot numbers. This viewpoint
survives even now among people not closely related to solar astronomy.
In fact, however, the solar activity is a complex of interconnected
physical processes mainly underlied by magnetic field variations.
The variety of physical processes included in solar activity is determined
by the fact that the solar magnetic field contains various components
and is produced by several drivers, such as, e.g., the differential rotation, 
meridional circulation, and the turbulent dynamo effects.
The differential rotation of the Sun is well known thanks to the helioseismology results (see, e.g. \citealp{Howe2011b}). The parameters of meridional circulation are debatable (see, e.g, the recent discussion in \citealp{Gizon2020} and \citealp{Stejko2021} and references therein). Our knowledge of the turbulent dynamo effects is even more uncertain \cite{Brandenburg2018}.  The spacial
and temporal magnetic field variations are important to understand
the very origin of the solar magnetic field as well as its relation
to the solar activity manifestations. Magnetic fields in various objects
of solar activity (sunspots, flares, filaments, corona, solar wind,
etc.) are addressed in many papers (see for review, e.g., \citealp{Hathaway2015},
\citealp{Usoskin2017LRSP}).

Here, we concentrate our analysis on the structure of the large-scale
global magnetic field presented as a combination of multi-poles of
various order. We are interested in the time evolution of individual
multi-poles as well as their correlations. The amplitudes and phases
of the lowest multipoles and their correlations with the solar photospheric
magnetic field have been investigated in various papers (e.g., \citealp{1977ApJ218291L,1984PhDT5H,1986Natur319285S,1987SoPh108205S,Hoeksema1991a,1977ApJ218291L,1992SoPh13835G,1992SoPh138399G,Stenflo1994,Stenflo2005})
on the basis of Kitt Peak and WSO (John Wilcox Observatory, Stanford)
data.

We will focus on higher-order axisymmetric multipoles in order
to determine the number $l$, which gives the main contribution to
the spectrum of the large-scale magnetic field. We believe that such
a research is interesting in itself, as well as in the context of
comparison with solar dynamo models. Now the most popular kind of
the solar dynamo, which successfully reproduces many observational
facts concerning the solar activity is the Babcock-Leighton scheme
(e.g., \citealp{Charbonneau2020}). 
 The polar magnetic field remaining from the previous cycle changes
its sing during the sunspot number growth phase (\citealp{Wang1989b},
\citealp{2004AA428L5B}, \citealp{2010AA518A7D}). 

Earlier, \cite{DeRosa2012} found  that the
coupling between the odd and even (anti-symmetric \& symmetric about
the equator) modes shows a correlation with the large-scale polarity reversals both in observations and in the Babcock-Leighton type dynamo models. The results obtained by \cite{Passos2014} and \cite{Hazra2014ApJ}, as well as the updated version of the  Babcock-Leighton dynamo model presented in \cite{Cameron17} show that the poloidal magnetic field can be generated in the solar dynamo by two different sources. One is related to the surface evolution of the tilted bipolar regions and the other comes from turbulent generation of the poloidal magnetic field from the large-scale toroidal field by the $\alpha$-effect in the bulk of the convection zone. The latter source is a usual component in the distributed mean-field dynamo models. Note that the distributed dynamo paradigm is supported by a number of the global convection simulation models (see, e.g., \citealp{Kapyla2016,Warnecke2018}). A comparison of the distributed dynamo model and the Babcock-Leghton scenario can be found in \citep{Brandenburg2005, Kosovichev2013}. 

{In this paper, we aim to investigate the relative contribution of  
the dynamo in the depth of the convection zone and the surface sunspot
activity to the evolution of the axisymmetric modes. The main idea is to study the phase relation between the evolution of different $\ell$ modes at the surface and to compare
the results of observations with the results of the dynamo models.
The phase relation between the mode evolution was previously studied
by \cite{Stenflo1994} and \cite{Stenflo2005}. They found an exceptional
role of the axisymmetric harmonic $\ell=5$. \cite{DeRosa2012} analyzed the
phase relations between the low $\ell$ harmonics in the Babcock-Leighton
type dynamo and revealed nearly synchronous evolution of the $\ell=1,3$
and $\ell=5$ (see Fig. 15 in \cite{DeRosa2012}). This seems to
contradict the results by \cite{Stenflo1994}. Here, we intend to repeat the analysis using observational data from the Wilcox Solar Observatory. }

.

\section{Observational basis and basic equations}

We have been working with the WSO (Stanford University) synoptic maps
of the light-of-sight photospheric magnetic field component (\citealp{Schetal77})
for the period from Carrington rotation 1642 (beginning on 27 May
1976) till rotation 2227 (beginning on 2 February 2020) converted
into Gaussian coefficients using the standard expressions 
\begin{eqnarray}
B_{r} & = & \sum_{l,m}P_{l}^{m}(\cos\theta)(g_{lm}\cos m\phi+h_{lm}\sin m\phi)\times\label{eq1}\\
 &  & \times((l+1)(R_{0}/R)^{l+2}-l(R/R_{s})^{l-1}c_{l}),\nonumber \\
B_{\theta} & = & -\sum_{l,m}\frac{{\partial P_{l}^{m}(\cos\theta)}}{{\partial\theta}}(g_{lm}\cos m\phi+\\
 &  & +h_{lm}\sin m\phi)((R_{0}/R)^{l+2}+(R/R_{s})^{l-1}c_{l}),\nonumber \\
B_{\phi} & = & -\sum_{l,m}\frac{m}{{\sin\theta}}P_{l}^{m}(\cos\theta)(h_{lm}\cos m\phi-\label{eq2}\\
 &  & -g_{lm}\sin m\phi)((R_{0}/r)^{l+2}+(R/R_{s})^{l-1}c_{l}).\nonumber 
\end{eqnarray}
Here, $0\le m\le l<N$, $c_{l}=-(R_{0}/R_{s})^{l+2}$, $R_{0}$, and
$R_{s}$ are the radii of the solar surface and the sours surface,
correspondingly; $P_{l}^{m}$ are the associated Legendre polynomials;
$g_{lm}$ and $h_{lm}$ are the Gauss coefficients calculated under
the assumption that the magnetic field is potential between the photosphere
and the source surface, while the magnetic field is presumed to be
exactly radial at the source surface. We do not include the correction
coefficient 1.8 suggested by \cite{Sv78}; however, it is more than
possible to re-scale the results below including this coefficient.
{The harmonic coefficients are generated on the base of the summary year-long synoptic map for every rotation with the cadence equal to half a Carrington rotation.}

Following \cite{Oetal20}, we obtain the Gauss coefficients $g_{lm}$
and $h_{lm}$ independently rather than just defining their ratio. Then, following \cite{OE89}, we average Eqs.~\ref{eq1} over the
photosphere and the source surface, respectively to yield
\begin{eqnarray}
<B_{r}^{2}>_{|_{R_{0}}}&=&\sum_{lm}\frac{{(l+1+l\zeta^{2l+1})}}{{2l+1}}(g_{lm}^{2}+h_{lm}^{2}), \label{eqav1} \\
<B_{r}^{2}>_{|_{R_{s}}}&=&\sum_{lm}(2l+1)\zeta^{2l+4}(g_{lm}^{2}+h_{lm}^{2}),
\label{eqav2} 
\end{eqnarray}
where $\zeta=R_{0}/R_{s}$ and $<...>$ stands for averaging. $<B_{r}^{2}>$
is the mean square radial component of the magnetic field calculated
by formula (1). Since the polynomials are orthonormal, the integration
can be done analytically. Integration can be performed over any spherical
surface by substituting the required value of R. In this case, the
formulas are given for the photosphere surface (4) and the source
surface (5). This means that the contribution of the $l$th mode to
the average magnetic field contains an l-dependent coefficient. Comparing
the relative contributions of modes with various $l$ we plot square
roots of contribution with given $l$ in Eqs.~(\ref{eqav1} - \ref{eqav2}).

Of course, we use for comparison more traditional solar activity data
like sunspots where required.

\section{Observational results: axial symmetry and parity  of the  $l$-harmonics}

\subsection{$l$-parity}

The solar magnetic field is traditionally claimed as a magnetic field
of dipole symmetry, i.e., antisymmetric in respect to the solar equator.
The most straightforward representation of such view is the famous
Hale polarity law of sunspot groups. Of course, the dipole symmetry
is not perfect and various deviations are known.

\begin{figure}
\centering \includegraphics[width=\columnwidth]{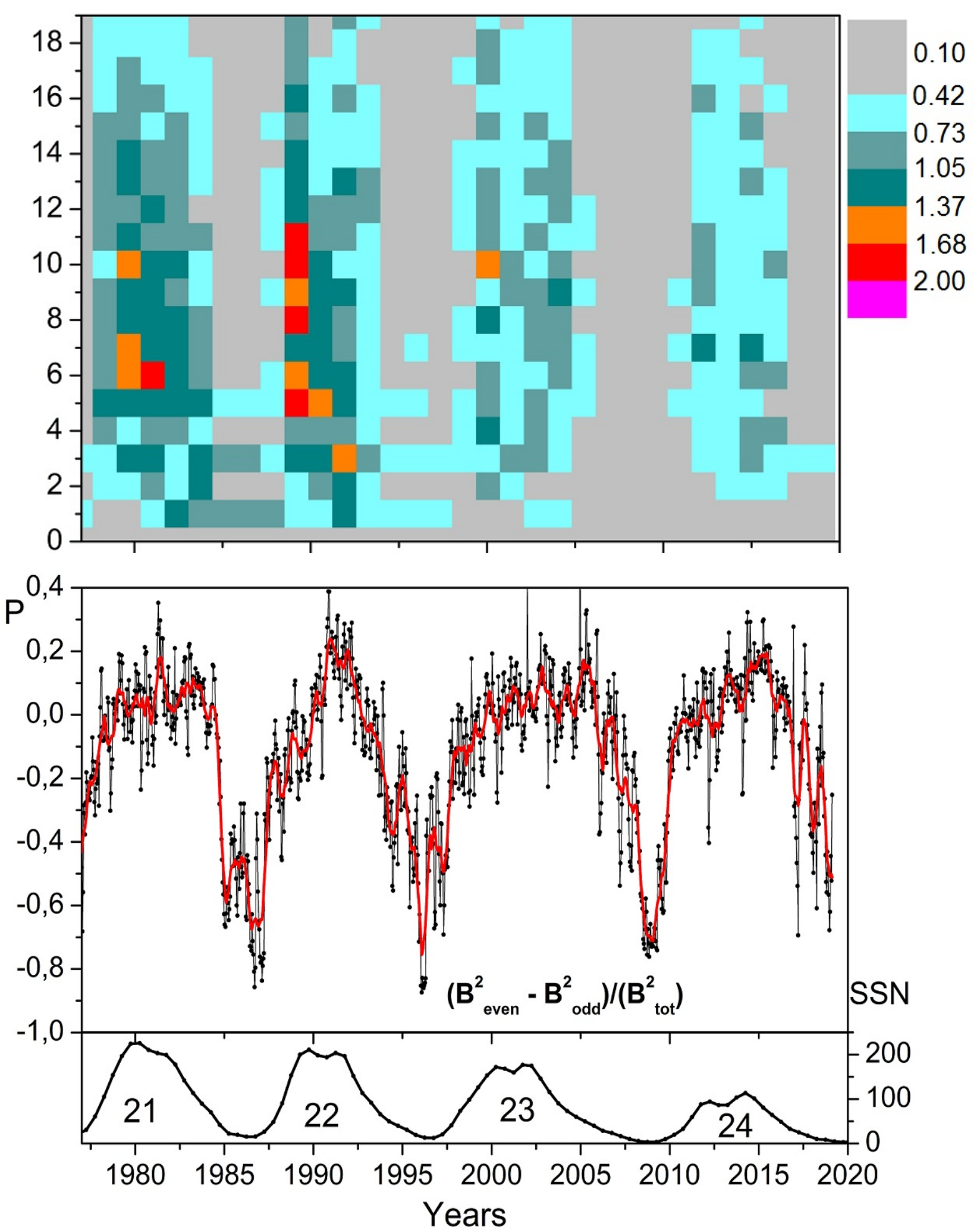} \caption{\label{Lplot}Top panel -- contribution of each of the first 20 harmonics
($l=0,1,2,...,19$) on the photosphere surface to the mean magnetic
field versus time. The middle  shows the parity index, P, where the black curves show the raw data; each dot corresponds to half a Carrington rotation, and the red curve represent 13-point smoothing. The bottom shows  the solar cycle according to sunspot data.}
\end{figure}

Fig.\ref{Lplot} (top panel) represents the contribution of various
$l$-harmonics to the mean surface magnetic field versus time. Some
prevalence of the even harmonics is visible. The integral parameter, which characterizes the equatorial symmetry of the large-scale magnetic field, is given by the parity index,  
\begin{equation}
P=(B_{{\rm even}}^{2}-B_{{\rm odd}}^{2})/(B_{{\rm even}}^{2}+B_{{\rm odd}}^{2}).    
\end{equation}
It reflects the relative power of the odd and even harmonics in the energy of the large-scale magnetic field. The negative values correspond to the equatorial anti-symmetry ($-1$
- the dipole parity) and the positive values, to the symmetry ($+1$
- the quadrupole parity). This parameter is commonly used in solar dynamo models \citep{Brandenburg1989,Knobloch1998,Weiss2016}.  From   Fig.\ref{Lplot} (middle), we see that the solar dynamo shows the dipole-type symmetry during the solar minimum. The solar maximum is characterized by the magnetic field distribution highly asymmetric about the equator.  Later, we shall discuss the origin of this phenomenon in more detail.

\subsection{Axisymmetric versus non-axisymmetric harmonics}

\begin{figure}
\centering \includegraphics[width=0.45\textwidth]{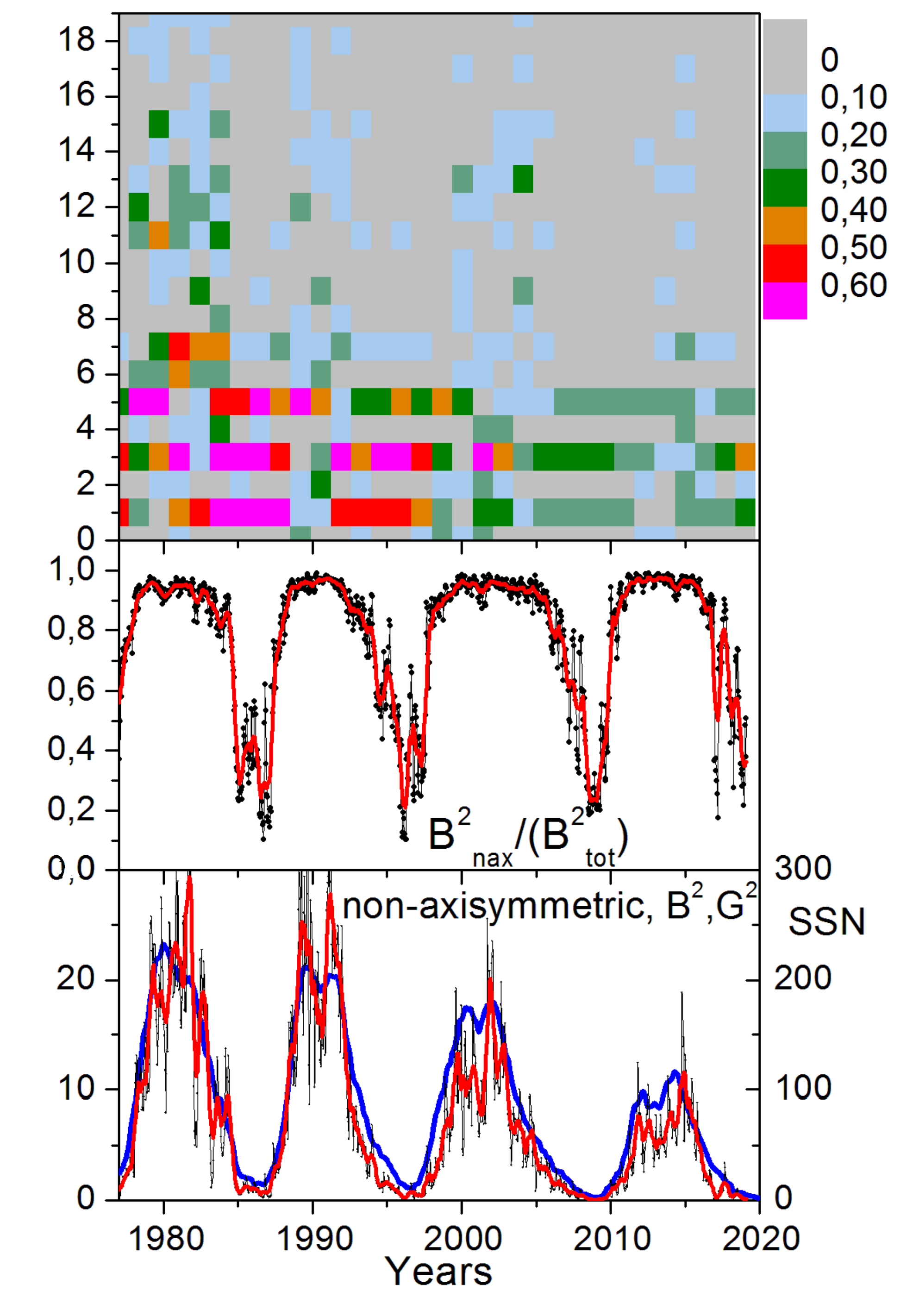} \caption{\label{2last} Contribution of each of the first 20 axisymmetric harmonics
($l=0,1,2,...,19$) to the mean magnetic field on the photosphere
surface versus time, $m=0$ (top); relative contribution of nonaxisymmetric
harmonics ($m\protect\ne0$) to the mean square magnetic field on
the photosphere surface versus time (middle); and total contribution
of nonaxisymmetric harmonics to the mean solar magnetic field (bottom).
The blue curve on the lower panel shows the solar cycle according
to sunspot data. }
\end{figure}

It is interesting that the evolution of the $\ell$ parity of the solar magnetic field seems to be related to the parameters of the axial symmetry. Fig.~\ref{2last} (top) shows the evolution of the $\ell$ power for the axisymmetric modes of the solar magnetic field during the past four solar cycles.  We see its substantial decreases at the end of the observational epoch. This indicates that solar activity has declined
substantially in the past two cycles as compared to the end of the
XX century. The integral parameter characterizing the relative power of nonaxisymmetric harmonics follows the evolution of the sunspot activity. This parameter is close to unity during the maximum of the sunspot activity (cf, \citealp{Vidotto2018}). The solar minima show a predominance of the  axisymmetric harmonics.

\begin{figure}
\includegraphics[width=\columnwidth]{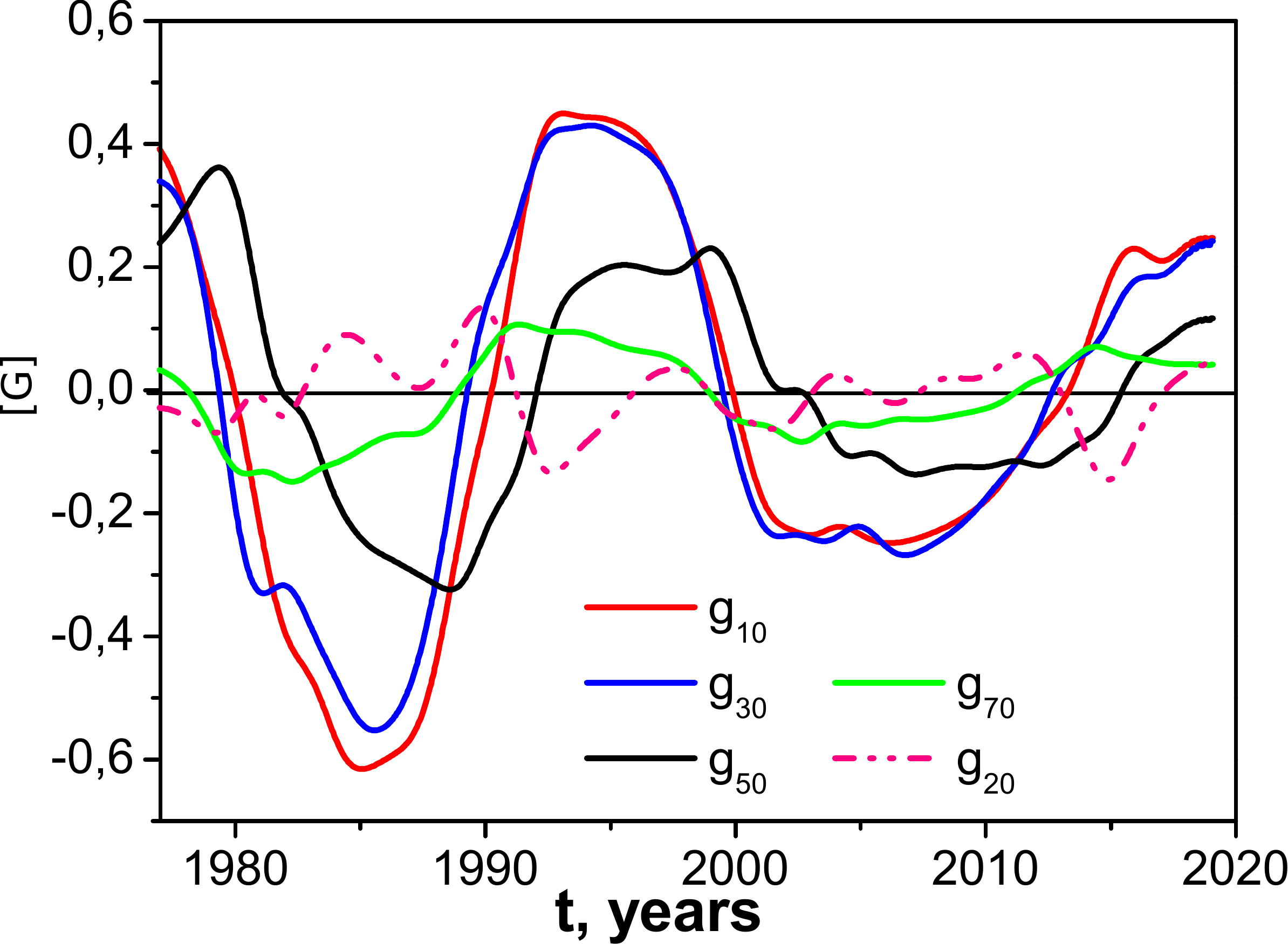} \caption{\label{modes}The smoothed series of the first 4 odd axisymmetric harmonics together with he lowest even axisymmetric harmonic. The
data averaged by convolution with the Gaussian function (parameter $\sigma=0.75$). }
\end{figure}

\section{Cyclicity in axisymmetric harmonics}

The analysis of the cyclic behavior of axisymmetric modes has run
into some difficulties.  As seen from Fig. 2, not only the even axisymmetric harmonics,
but also the higher-order odd ones yield a chaotic, unrealistic picture.
Apparently, taking into account any harmonics of the order higher
than $l=7$ results in violation of the regular cyclic dependence
(see, \citealp{hoeks-polar}, http://wso.stanford.edu/Polar.html).

\begin{table}
\caption{ \label{Tabl1} Phase shifts of the first axisymmetric harmonics with
the odd numbers juxtaposed with the phase of the solar cycle as inferred
from the polar magnetic field and sunspot data. We have chosen the phase of $g_{50}=0$.
The correlation functions (except the one for sunspots) have two maxima and, correspondingly, two estimates of the phase shift. The estimate corresponding to the
main maximum of the correlation function is given as the basic value
and the other value is added in brackets. The discrepancy between
the main quantities and the estimates given in brackets provide us
with additional option for measuring the accuracy of the estimates. }

\centering{}%
\begin{tabular}{|c|c|c|}
\hline 
 & $\Delta t$, years  & $\phi$\tabularnewline
\hline 
$g_{10}$  & $-2.28$ ($-2.03$)  & $0.23\pi$ ($0.20\pi$) \tabularnewline
\hline 
$g_{30}$  & $-2.69$ ($-2.29$)  & $0.27\pi$ ($0.23\pi$) \tabularnewline
\hline 
$g_{50}$  & 0  & 0 \tabularnewline
\hline 
$g_{70}$  & $-3.69$ ($-3.11$)  & $0.37\pi$ ($0.31\pi$) \tabularnewline
\hline 
$B_{{\rm polar}}$  & $-1.91$ ($-0.97$)  & $0.19\pi$ ($0.10\pi$)\tabularnewline
\hline 
Sunspot number  & $3.29$  & $-0.33\pi$ \tabularnewline
\end{tabular}
\end{table}
\begin{figure}
\includegraphics[width=0.99\columnwidth]{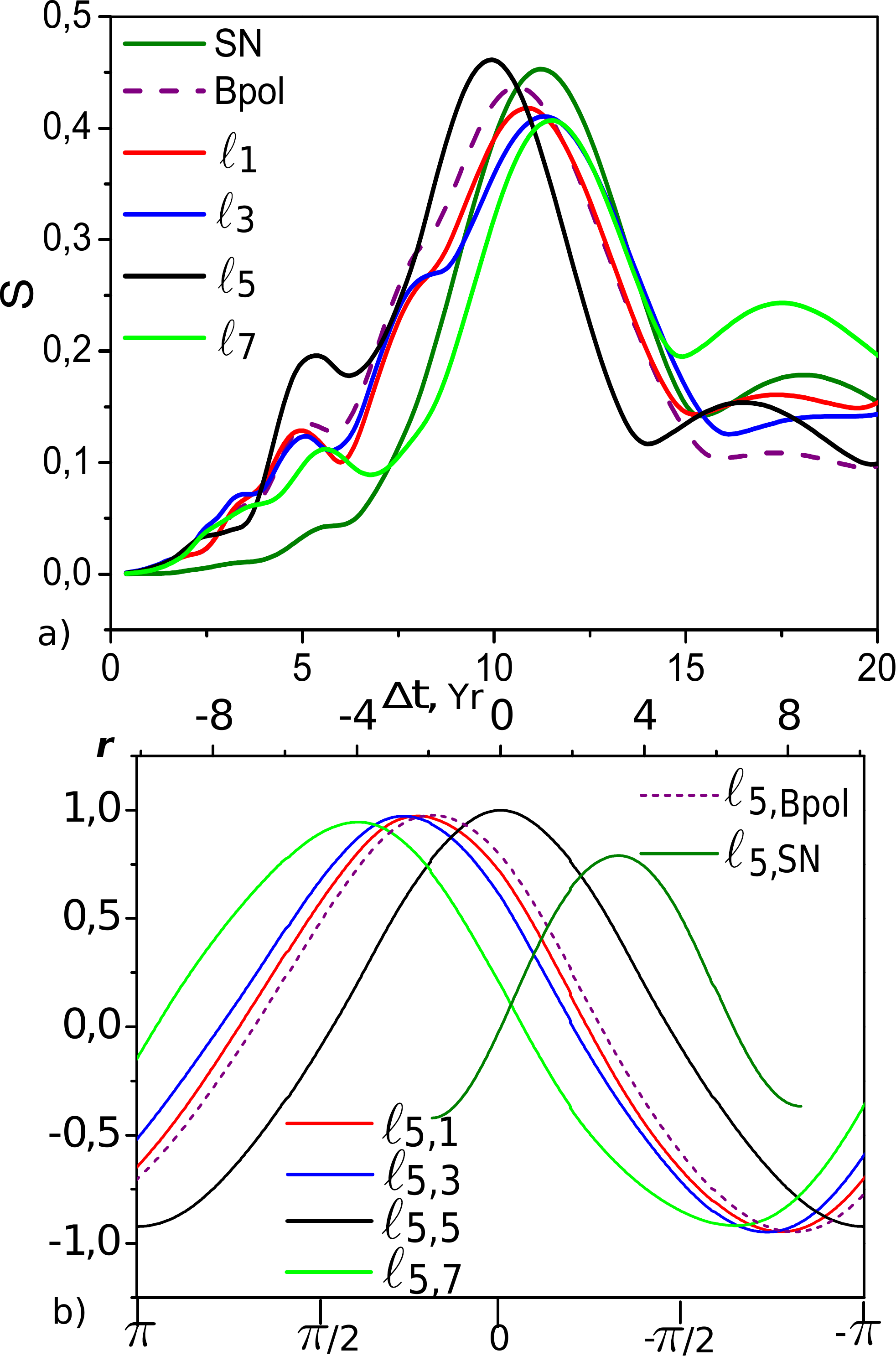} \caption{\label{Cor}a)
Integral wavelet spectra for several odd harmonics, the strength of the polar magnetic field and the SN index; b) Correlation coefficients $r$ between $g_{50}$ and the
other tracers of solar activity under discussion ($g_{10}$ - red,
$g_{30}$ - blue, $g_{50}$ - black, and $g_{70}$ - light green)
in comparison with the polar field $B_{{\rm polar}}$ (purple dashed)
and sunspots (dark green). The phase of the cycle is given in years
(upper horizontal scale) in phase $\phi$ of the whole cycle, which
is $2\pi$. }
 
\end{figure}

Then, we search for harmonics with a pronounced cyclic behavior and
select four such harmonics, namely, $g_{10}$, $g_{30}$, $g_{50}$, 
and $g_{70}$, Fig.~\ref{modes}. We do not see pronounced cyclicity
in axisymmetric harmonics with the even numbers; however, they are
interesting as indicators of the North-South asymmetry \citealp{DeRosa2012}. To be specific,
the red dashed line in Fig.~\ref{modes} show the lowest harmonic $g_{20}$.
Despite the activity of $g_{20}$ is irregular it shows  a correlation with the polar
field reversals \citep{DeRosa2012}. 
Fig.~\ref{modes} shows a phase difference in the evolution of the odd axisymmetric harmonics. This difference was analyzed by \cite{Stenflo1994}. He found, in particular, that an extended solar cycle in the evolution of the radial magnetic field  can inherently result from the evolution of modes, which follows some phase relations. Therefore, the phase relations can shed light on the origin of the solar dynamo. The analysis of  \cite{Stenflo1994} was based on the assumption of a unique period for  the axisymmetric modes. Here we consider a more general situation.

Fig.~\ref{Cor}a shows the integral wavelet spectra for the low odd harmonics, the strength of the polar magnetic field and the SN index. 
The cycle lengths of the modes, as well as the corresponding quantities
for the polar filed and the sunspot numbers determined as the location
of the maxims in the integral wavelet spectra. They are given in Table~\ref{Tabl2}.
One can see, that the different parameters show the different cycle
length. {Notably, The parameter $g_{50}$ shows the smallest period of
about 10 years. }

In order to obtain the phase shift between the axisymmetric harmonics
and some other tracers of solar activity, we calculate the corresponding
correlation coefficients choosing quite arbitrarily the harmonic $g_{50}$
as the basic one. More specifically, we consider two signals, say,
$f(t)$ and $g(t)$, shift signal $g$ in time by $\Delta t$, and
calculate the correlation function $r$. The result is plotted in
Fig.~\ref{Cor}b as a function of $\Delta t$. 
{For comparison with \cite{Stenflo1994}, we have chosen the phase of $g_{50}$ equal to $0$.
Therefore, the mode $g_{50}$ in Fig~\ref{Cor}b is shown by the auto-correlation coefficient. In order to compare with the above-cited paper, we have to take into account that the highest correlation is 
attained by shifting the advance harmonic backward or by shifting the lagged harmonic forward. Therefore, the signs for the time and phase are opposite. Similarly to  \cite{Stenflo1994} we find the positive sign for the phase shifts of the all harmonics. Though the magnitude of the shift is about factor twice less than in his study.}  Recall that
$r$ is a dimensionless quantity and $|r|\le1$. The location of maximum
$r$ for the particular tracer of the solar activity is used as the
phase shift of the tracer in respect to the mode $g_{50}$. The autocorrelation
coefficient for $g_{50}$ is also shown in the figure for comparison.

Note that the maximum values of $r$ are quite high exceeding 0.9
for $g_{10}$ and $g_{30}$, and even the lowest value (obtained for
sunspots) exceeds 0.7. The fact that correlations between the harmonics
of the large-scale poloidal magnetic field are larger than their correlations
with the polar magnetic field and the correlations with sunspots are
even lower looks reasonable because it agrees with the degree of connections
between these quantities as expected from the solar dynamo mechanism. 

Based on the Fig.~\ref{Cor}b, we calculate the corresponding phase
shifts as location of the maxima in $\Delta t$ (presented in Table~1).
The width of the plot for a particular correlation coefficient allows
us to decide how the main phase of the solar cycle is localized for
the particular harmonics. In particular, it looks difficult to insist
that the phase shifts between $g_{10}$, $g_{30}$, and $B_{{\rm polar}}$
are significant; however, the phase shifts between them and, perhaps,
$g_{70}$, $g_{50}$ and sunspots number seem to be significant. {It follows from Figure 5b and Table 1 that the sunspot cycle is ahead of the polar field cycle by $\pi/2$, and the $g_{50}$ mode, by  $\pi/3$.}

\begin{table}
\caption{ \label{Tabl2} Cycle lengths of the modes under discussion, polar
magnetic field, and sunspot numbers: $T$ is the cycle length for
the absolute value, $T_{s}$ is the cycle length for the signed quantities. }

\centering{}%
\begin{tabular}{|c|c|c|}
\hline 
 & $T$ (years)  & $T_{s}$ (years) \tabularnewline
\hline 
$g_{10}$  & 11.5  & 23.0 \tabularnewline
\hline 
$g_{30}$  & 11.75  & 23.5 \tabularnewline
\hline 
$g_{50}$  & 9.75  & 19.5 \tabularnewline
\hline 
$g_{70}$  & 11.625  & 23.25 \tabularnewline
\hline 
$B_{{\rm polar}}$  & 10.5  & 21.0 \tabularnewline
\hline 
Sunspot number  & 11.0  & 22.0 \tabularnewline
\end{tabular}
\end{table}

The analysis of observations can be summarized as follows.  The actual phase and
length of a nominal 11-year cycle as recorded in such tracers are
specific for each harmonic. {This was already demonstrated
in \cite{1986Natur319285S},  \cite{Stenflo1994} and
\cite{Stenflo2005}. Compared to their work, our analysis
covers the period of two full magnetic cycles. Therefore, our conclusions
concerning the phases and frequencies of the large-scale magnetic field
spherical harmonics are more robust. The low $\ell$ axisymmetric
harmonics show a phase shift relative to each other and relative to the proxy
of the polar magnetic field strength and the sunspot number index.
We have to note that the phase shifts mostly exceed the
difference between the mode oscillation periods at least
by factor of two. This observation may tell us about the spatial distribution
of turbulent sources in the solar dynamo. Indeed, the results of \cite{DeRosa2012}
regarding the phase relations between the low $\ell$ harmonics in the Babcock-Leighton
type dynamo model show a nearly synchronous evolution of $\ell=1,3$
and $\ell=5$ (see Fig. 15 in \cite{DeRosa2012}). This doesn't seem
surprising, because in their model, the sources of generation of the radial
magnetic field  are in one place, near the solar surface.
Below, we consider the results of the distributed dynamo model that
include the effects of emergence of a bipolar region in order to mimic the effects
of the sunspot activity on the evolution of the radial magnetic field.}

\section{Results from mean-field dynamo models}

\begin{figure*}
\includegraphics[width=0.75\textwidth]{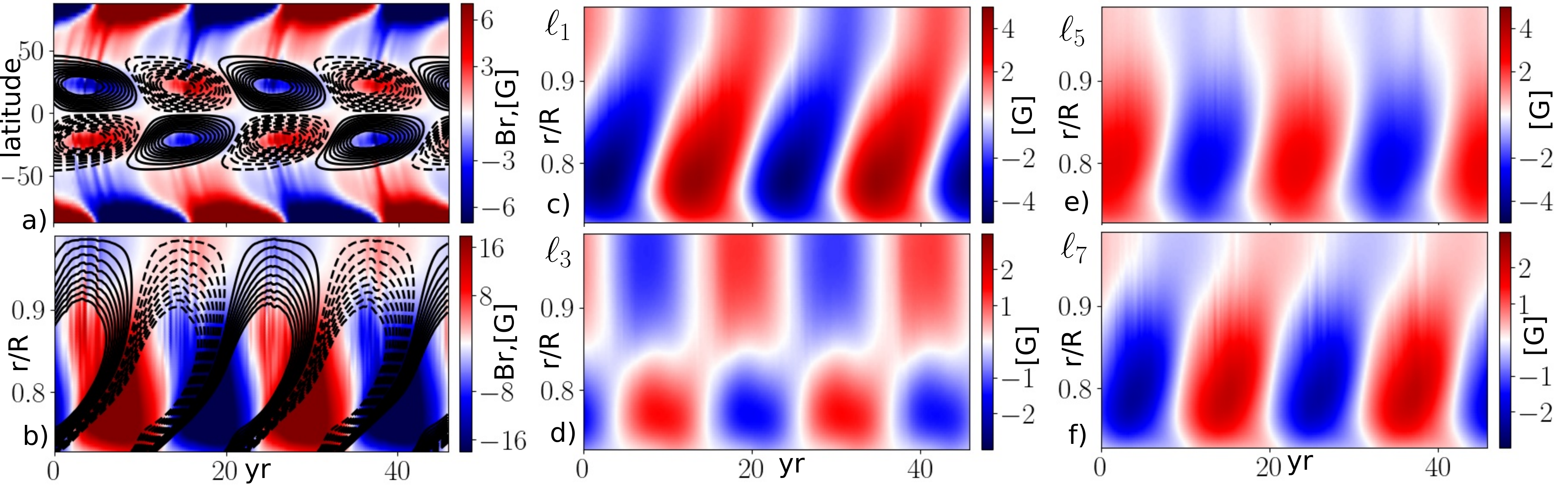} \caption{\label{fig:nxd}a) The time-latitude diagrams of the toroidal magnetic
field (contours in the range of $\pm$1kG) in the subsurface shear
layer, r=0.9R and the radial magnetic field on the surface (background
image); b) the time-radius diagram for the large-scale magnetic field
at the latitude of 30$^{\circ}$; c) the time-radius evolution of
the axisymmetric mode of the radial magnetic field $\ell_{1}$; d),
e), and f) show the same for $\ell_{3}$, $\ell_{5}$, and $\ell_{7}$.}
\end{figure*}

\begin{figure*}
\includegraphics[width=0.75\textwidth]{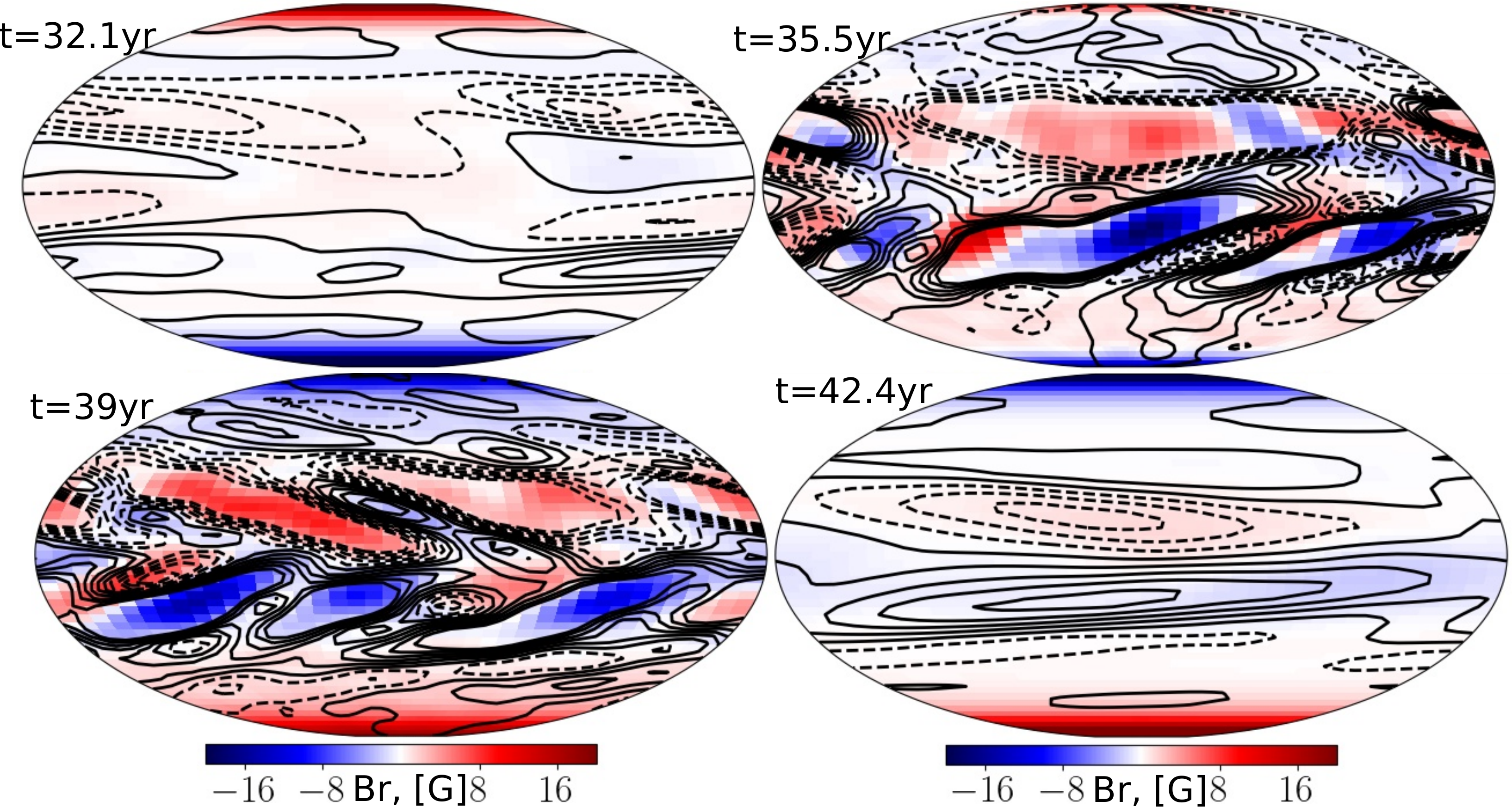} \caption{\label{snps} Snapshots of the surface magnetic field distributions
in a dynamo cycle. The radial magnetic field is shown by color and
the toroidal magnetic field, by contours in the same range of $\pm20$G.}
\end{figure*}

\begin{figure}
\includegraphics[width=0.9\columnwidth]{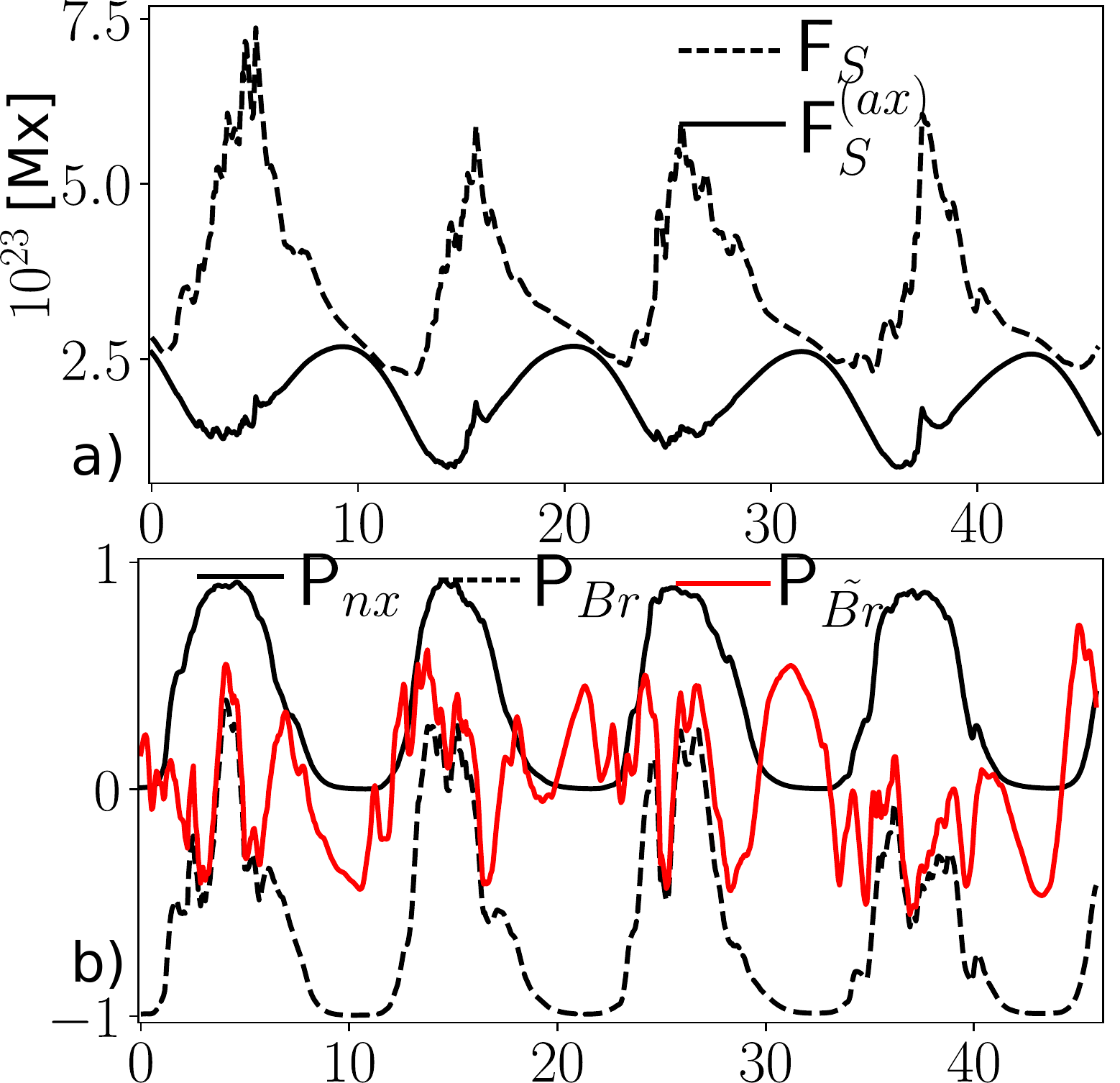}

\caption{\label{fig:int}a) $F_{S}$ is the total unsigned flux of the radial
magnetic field and $F_{S}^{(ax)}$ is the same for the contribution
of the axisymmetric magnetic field; b) $P_{nx}$ is the parameter
characterizing the nonaxisymmetry of the surface radial magnetic field,
$P_{Br}$ is the equatorial parity parameter of the surface radial
magnetic field, and $P_{\widetilde{Br}}$ is the same for the nonaxisymmetric
radial magnetic field.}
\end{figure}

In a preliminary study we considered the results of the run
C2 from the axisymmetric dynamo model of \cite{Pipin2020}. A detailed description of the model is given in the above-cited paper. Note that the model explains successfully the so-called ``extended''  solar cycle (\cite{Altrock1997}) and, besides that, it shows a satisfactory agreement with many aspects of solar observations. However, that model does not take into account the effect of the sunspot activity on the evolution of the radial
magnetic field. Despite this, it shows a satisfactory
agreement of evolution of the low $\ell$ modes with observations.
We provide these results in Appendix B. This model will be a reference in our study of the relative contributions of the deep and surface dynamo processes to the whole magnetic variability of the Sun. To study the possible effects of the sunspot activity on the dynamo evolution, we consider the nonaxisymmetric dynamo model. 

We assume that the toroidal magnetic field in the subsurface layer
of the convection zone is responsible for the formation of solar bipolar
regions in the photosphere. The effect of inclination of the bipolar
regions can be parametrized by means of an additional $\alpha$-effect
acting on the emerging part of the toroidal magnetic field. This idea
follows the phenomenological approach of the Babcock-Leighton dynamo
scenario. The mean-field formulation of this scenario was
suggested by \cite{Nandy2001}. Here, we follow the same ideas (in
particular, see  \cite{Kumar2019}). In this scenario,
the individual magnetic flux-tubes that rise from the bottom of the
convection zone are subject to the effect of the Coriolis force (cf,
\citealp{Cameron17}). In the dynamo equations, the averaged effect
of this process is described by means of the $\alpha$-effect acting
on the axisymmetric magnetic field. Unlike the other parts of the
mean electromotive force, the analytical expression of this effect
was never obtained from the first principles. In our model we do not
apply this ansatz and consider the $\alpha$-effect acting on the
buoyant nonaxisymmetric parts of the large-scale toroidal magnetic
field. The description of this $\alpha$-effect in our model
remains speculative and has a phenomenological character. We simulate
the emergence process by means of the magnetic buoyancy of the randomly
chosen parts of the axisymmetric toroidal magnetic field in the upper
part of the convection zone.

The general consideration shows that the large-scale toroidal magnetic
field can be unstable if its strength decreases outward faster than
the mean density stratification does \citep{Gilman1970}. The instability
is subjected to effects of the global rotation and turbulent diffusion
\citep{Gilman1970,Acheson1978,Hughes2011,Gilman2018}. In our model,
these effects are not taken into account. The nonaxisymmetric magnetic
buoyancy and the related $\alpha$-effect are used in a phenomenological
way to generate a solar-like large-scale nonaxisymmetric magnetic
field at the dynamo surface. The current knowledge shows that the
magnetic buoyancy is unlikely to be the only process responsible for
the sunspot formation. All complicated phenomena, such as the effects
of convective flows, instabilities due to interaction of the magnetic
field and turbulent plasma, and magnetic buoyancy instability, should
be taken into account \citep{Kitiashvili2010a,Stein2012,Losada2017}.
In our model we keep the consideration as simple as possible. Also,
we do not try to resolve the real spatial scale of the sunspots active
regions. This is rather expensive computationally. The 2D analog of
the model was recently discussed by \cite{Pipin2020b}. The model
should be considered as a numerical experiment. It does not pretend
to explain the emergence of the solar active regions. The further
details concerning the nonaxisymmetric dynamo model are given in the
Appendix A.

Figure \ref{fig:nxd} shows the time-latitude and the time-radius
diagrams of the large-scale magnetic field evolution, as well the
time-radius evolution of the first four odd axisymmetric spherical
harmonics in the nonaxisymmetric dynamo model. The results for the
axisymmetric modes $\ell_{1-7}$ are qualitatively similar to the
results of the axisymmetric model. However, the time-radius evolution
of the radial magnetic field at the latitude of 30$^{\circ}$ shows
the difference. The radial magnetic field generated in the subsurface
shear layer drifts downward from the surface and has the opposite
sign to the magnetic field propagating from the depth of the convection
zone. The poleward surges in the time-latitude diagram of the large-scale
radial magnetic field are due to the formation of bipolar regions.
There are similar radial surges in the time-radius diagrams.

Figure \ref{snps} illustrates the snapshots of the surface magnetic
field distributions during a dynamo cycle starting from the minimum
of the magnetic activity. The model shows solar-like patterns of the
magnetic field distribution during the growing and maximum phases
of the magnetic activity. Comparing these results with \cite{Vidotto2018},
we find that the magnetic field distribution patterns agree with the
low-resolution patterns of the magnetic activity found in observations.

The magnitude of the nonaxisymmetric magnetic field in these snapshots
is rather small compared to the sunspot magnetic field. This is due
to our model restrictions on the strength and size of the bipolar
region. However, the amplitude of the generated radial magnetic field
flux agrees with observations to the order of magnitude. The total
flux of the radial magnetic field in the model is about 6~10$^{23}$
Mx (see Figure\ref{fig:int}a). This is lower than the value reported
by \cite{Schrijver1994} (10$^{24}$Mx) and higher than the flux obtained
in the WSO low-resolution observations. The total magnetic flux of
the radial magnetic field in the model evolves in phase with the toroidal
magnetic field activity in the subsurface shear layer. The magnetic
flux of the axisymmetric magnetic field is shifted by $\pi/2$ relative
to the evolution of the axisymmetric toroidal magnetic field. Using
decomposition of the radial magnetic field into the partial dynamo
modes: 
\[
\left\langle \mathbf{B}_{r}\right\rangle =\sum B_{r}^{(m)}\left(\mu\right)\mathrm{e^{im\phi}},
\]
where the case $m=0$ corresponds to the axisymmetric magnetic field,
we compute the degree of nonaxisymmetry of the magnetic field: 
\begin{eqnarray}
P_{nx} & = & {\displaystyle \frac{\tilde{E}_{r}^{(m)}}{E_{r}^{(m)}},}\label{eq:nx}
\end{eqnarray}
where 
\begin{equation}
\tilde{E}_{r}^{(m)}=\frac{1}{8\pi}\sum_{m\ge1}\intop B_{r}^{(m)}B_{r}^{(m)*}d\mu,\label{eq:rm}
\end{equation}
and $E_{r}^{(m)}$ is the total energy of the radial component of
the magnetic field. Using decomposition of the radial magnetic field
into spherical harmonics, we compute the two parity indices: 
\begin{equation}
P_{Br}=\frac{\sum B_{r}^{(\ell_{{\rm even}})2}-\sum B_{r}^{(\ell_{{\rm odd}})2}}{\sum B_{r}^{(\ell)2}},\label{eq:pr}
\end{equation}
where summation is done for all modes, and a similar one for the nonaxisymmetric
modes:

\begin{equation}
P_{\widetilde{Br}}=\frac{\sum^{m>0}B_{r}^{(\ell_{{\rm even}})2}-\sum^{m>0}B_{r}^{(\ell_{{\rm odd}})2}}{\sum^{m>0}B_{r}^{(\ell)2}}.\label{eq:prnx}
\end{equation}
Figure \ref{fig:int}b shows the evolution of these parameters in
the nonaxisymmetric dynamo model. In general we see a qualitative
agreement of our results with the observation data represented in
Figure \ref{2last}. In this phase of the magnetic cycle, the parameter
$P_{nx}$ is minimum, i.e., the energy of the nonaxisymmetric magnetic
field reaches the lowest value.

The model shows substantial variations in the parity parameters of the axisymmetric, $P_{Br}$, and non-axisymmetric, $P_{\widetilde{Br}}$, radial magnetic field. In the model, the emergence of the bipolar region is random in longitude and hemisphere. This induces the hemispheric asymmetry of the sources  of the radial magnetic field. In our model the parity varies from -1 to the values which are slightly above zero level. The zero parity means a strong hemispheric asymmetry of the magnetic activity. The obtained parity variations agree with the results of  observations shown in Fig.1.  The results of our model are qualitatively similar to those demonstrated by the 3D Babcock-Leighton type model of \cite{Hazra2019n}. When comparing the results of the two dynamo models, we have to take into account the difference in the parity parameter definitions. Our definition of the parity parameter reflects the difference between the energy of a symmetric and antisymmetric magnetic field about the equator (see, \citealp{Brandenburg1989,Knobloch1998}).  Also, our model covers the period of only a few magnetic cycles. The  dynamo simulations of \citep{Moss2008,Weiss2016, Mordvinov2018, Hazra2019n} display much stronger long-term parity variations. The above cited papers show that these variations can result both from fluctuations of the $\alpha$ effect and from hemispheric randomness in the formation of a bipolar region. Such long-term variations are beyond the scope of this paper.

\begin{figure}
\includegraphics[width=0.85\columnwidth]{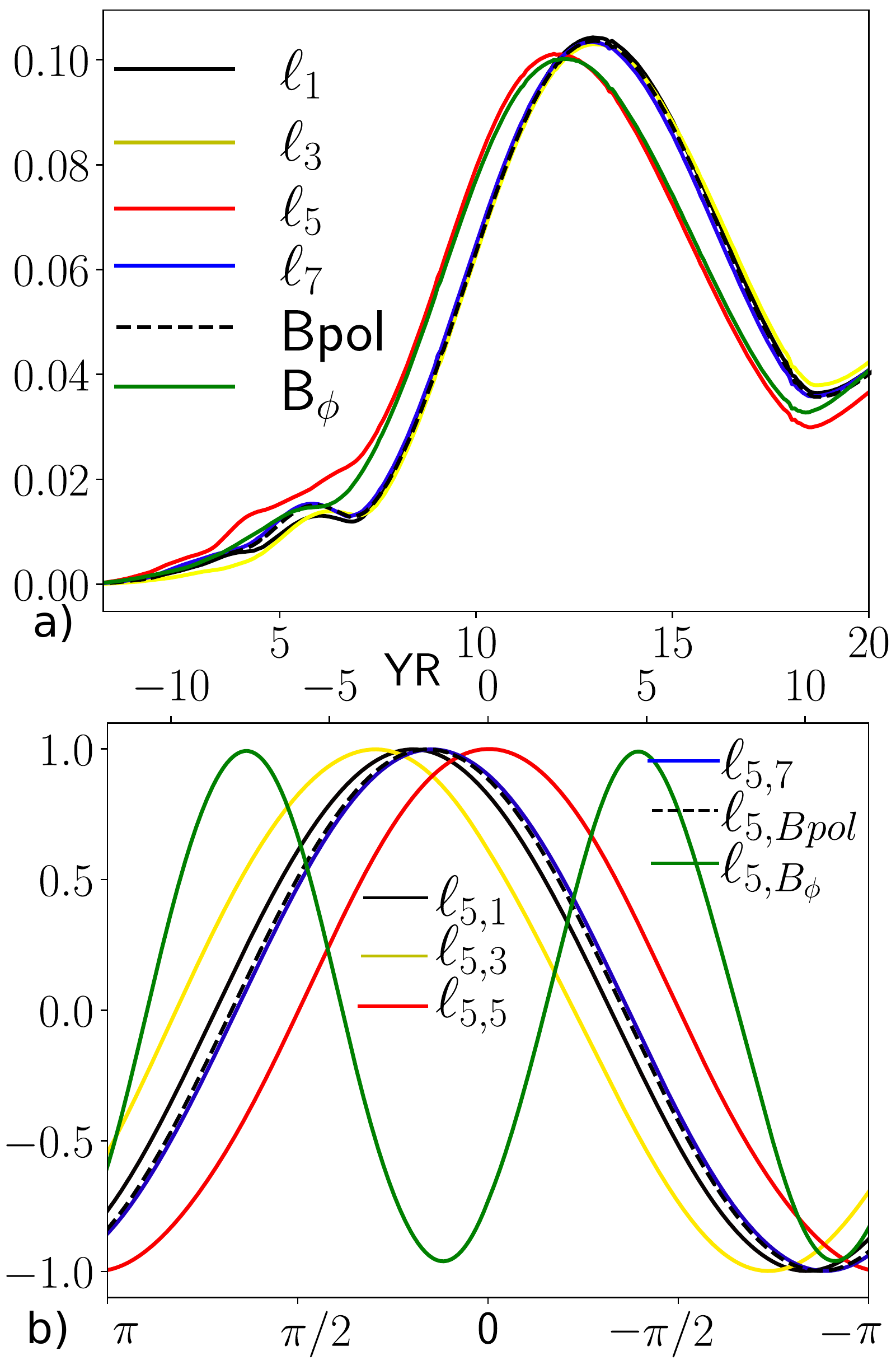} \caption{\label{CfM3}The same as Figures \ref{Cor} for the nonaxisymmetric dynamo model.}
\end{figure}

Figure \ref{CfM3} shows the integral wavelets of the low axisymmetric
modes and their cross-correlations with the $\ell_{5}$ mode for the
nonaxisymmetric dynamo model. The obtained results agree with observations
illustrated in Figures \ref{Cor}, except for the phase shift between
the modes $\ell_{5}$ and $\ell_{7}$. We find that the effect of
nonaxisymmetric magnetic fields brings the correlation between $\ell_{5}$
and the strength of the toroidal magnetic field in agreement with
observations. 
\begin{table}
\caption{\label{T-mod} Phase shifts (years) for the run C2 (see, Pipin\&Kosovichev,
2020) and the nonaxisymmetric dynamo model (NXDY).}

\begin{tabular}{ccc|cc}
\hline 
 & C2  & Period  & \multicolumn{2}{c}{NXDY,~ Period}\tabularnewline
\hline 
\multirow{2}{*}{$g_{10}$} & -2.  & \multirow{2}{*}{11.3} & -2.5  & \multirow{2}{*}{10.8}\tabularnewline
\cline{2-2} \cline{4-4} 
 & 0.17$\pi$  &  & 0.23$\pi$  & \tabularnewline
\hline 
\multirow{2}{*}{$g_{30}$} & -3.45  & \multirow{2}{*}{10.7} & -3.9  & \multirow{2}{*}{11.3}\tabularnewline
\cline{2-2} \cline{4-4} 
 & 0.3$\pi$  &  & 0.37$\pi$  & \tabularnewline
\hline 
\multirow{1}{*}{$g_{50}$} &  & 11.3  & 0  & 10.6\tabularnewline
\hline 
\multirow{2}{*}{$g_{70}$} & 1.1  & \multirow{2}{*}{11.3} & 1.81  & \multirow{2}{*}{10.7}\tabularnewline
\cline{2-2} \cline{4-4} 
 & 0.08$\pi$  &  & 0.18$\pi$  & \tabularnewline
\hline 
\multirow{2}{*}{$B_{{\rm pol}}$} & -1.8  & \multirow{2}{*}{11.3} & -2.1  & \multirow{2}{*}{10.8}\tabularnewline
\cline{2-2} \cline{4-4} 
 & 0.15$\pi$  &  & 0.21$\pi$  & \tabularnewline
\hline 
\multirow{2}{*}{$B_{\phi}$} & 3.95  & \multirow{2}{*}{10.6} & 3.1  & \multirow{2}{*}{10.8}\tabularnewline
\cline{2-2} \cline{4-4} 
 & -0.35$\pi$  &  & -0.29$\pi$  & \tabularnewline
\end{tabular}
\end{table}

The Table \ref{T-mod} provides the phase shifts and the dynamo periods
of the low axisymmetric modes and the integral parameters for our
dynamo models.

\section{Conclusion and Discussion}

Let us summarize the main results of our study. {Observations show the difference in the oscillation frequencies of the large-scale magnetic field spherical harmonics. Also, the low $\ell$ axisymmetric harmonics display a phase shift relative to each other and relative to the proxy of the polar magnetic field strength and the sunspot number index. Earlier this phase shift was discussed by \cite{Stenflo1994}. Our findings show a a factor twice less shift's magnitudes. This  is likely due to the difference of the time-series interval and the method of analysis. Indeed, the time-series in Fig3 seems to show a twice larger shifts which is in agreement with results of \cite{Stenflo1994} for period before 1990 year. Therefore the times shifts between the $\ell$ harmonics are not constant.}

Combining results of this and previous study of  \cite{DeRosa2012} we see that the phase shift between the modes depends on the spatial sources of the dynamo processes. For example,  the dynamo model by \cite{DeRosa2012} shows nearly synchronous oscillations with the zero phase shift. Our models are based on the distributed dynamo, and they show  different phases for the low $\ell$ modes. Both observations and the dynamo model show an exceptional role of
the axisymmetric $\ell_{5}$ mode. Its origin seems to be readily connected with the formation and evolution of sunspots on the solar surface. The fluctuating nature of this process can result in a dispersion of the dynamo periods. 
We find that the cycle periods and  the phase shifts of the low $\ell$ modes are different in the axisymmetric and nonaxisymmetric models. Thus, the asynchronous and shifted cycles of the mode can be accounted for both by the specific spatial distribution of the dynamo sources and by nonlinearities involved in the dynamo processes. 

In our model, the surface poloidal magnetic field originates from two sources: the poloidal magnetic field, which propagates with the dynamo wave from the depth of the convection zone, and the solar active regions that, in turn, are associated with the
dynamo driven toroidal magnetic field in the subsurface shear layer. This is similar to  studies of \cite{Passos2014}, \cite{Hazra2014ApJ} and  \cite{Hazra2019n}. Their models employed the Babcock-Leighton type scenario as the main dynamo process. The  need for a weak background distributed dynamo was justified by the temporal parameters of the magnetic activity during the Maunder minimum. The relative contribution of both sources is not known in advance and may be specific for each harmonic. The polar magnetic field seems to be most directly connected with the propagation of the poloidal magnetic field from the domain of the solar dynamo action, just because sunspots and active regions are located at middle solar latitudes.

Another interesting observational phenomenon is the excitation of even modes during the maximum phases of the sunspot activity. This results in variation of the parity index $P_{Br}$ around zero (see, Figs.1 and 4). This effect is readily connected with the hemispheric asymmetry of the magnetic activity. It was extensively discussed by \cite{DeRosa2012}  and \cite{Hazra2019n} as the source of long-term variations of the magnetic activity. The models of the Babcock-Leighton type show that the stochastic emergence of active regions and the induced stochastic variations of the $\alpha$-effect can explain both the long-term variations of the solar activity and variations in the hemispheric asymmetry of the activity
\citep{Mordvinov2018,Bhowmik2018NatCo, Nepomnyashchikh2019,Hazra2019n}. Note, that beating between the modes, when they have close but different periods, is often considered as a source of long-term variation of the solar magnetic activity
\citep{Brandenburg1989,Sokoloff1994,Knobloch1998,Feynman2014, Weiss2016,Beer18}. The random fluctuations of the dynamo governing parameters are usually considered as the main drivers of long-term variations of the sunspot activity (see the above cited papers).
The long-term parity variations cause a phase-shift between the magnetic activity in the North and South solar hemispheres (e.g., \citealp{Shib16,Beer18,Hazra2019n}). 
These long-term variations are accompanied by variations in the magnitude and period of a sunspot cycle  \citep{Hathaway2015}. It is likely, that stochastic fluctuations of the dynamo parameters can drive the phase shifts and frequencies of the individual $\ell$ harmonics as well.

\textbf{Acknowledgements}

The authors are grateful to the WSO teams for a free access to their
data. The work was supported by RFBR grants
No. 20-02-00150,19-52-53045. DS was supported by BASIS found number 18-1-1-77-3.
VVP thanks the financial support of  the Ministry of Science and Higher Education of the Russian Federation (Subsidy No.075-GZ/C3569/278).

Data Availability Statements. The data underlying this article are
available at \cite{SIDC2010}. The data of the nonaxisymmetric dynamo model are available at zenodo, \cite{Pipin2021d}. 

\bibliographystyle{mnras}
\input{lharmv2.bbl}

\section{Appendix}

\subsection*{A. Nonaxisymmetric dynamo model\label{supp}}

Evolution of the large-scale magnetic field in perfectly conductive
media is described by the mean-field induction equation \cite{Krause1980}:
\begin{equation}
\partial_{t}\left\langle \mathbf{B}\right\rangle =\mathbf{\nabla}\times\left(\mathbf{\mathbf{\mathbf{\mathcal{E}}}+}\left\langle \mathbf{U}\right\rangle \times\left\langle \mathbf{B}\right\rangle \right)\,,\label{eq:mfe}
\end{equation}
where $\mathbf{\mathcal{E}}=\left\langle \mathbf{u\times b}\right\rangle $
is the mean electromotive force; $\mathbf{u}$ and $\mathbf{b}$ are
the turbulent fluctuating velocity and magnetic field, respectively;
and $\left\langle \mathbf{U}\right\rangle $ and $\left\langle \mathbf{B}\right\rangle $
are the mean velocity and magnetic field. It is convenient to represent
the vector $\left\langle \mathbf{B}\right\rangle $ in terms of the
axisymmetric and non-axisymmetric components as follows: 
\begin{eqnarray}
\left\langle \mathbf{B}\right\rangle  & = & \overline{\mathbf{B}}+\tilde{\mathbf{B}}\,,\label{eq:b0}\\
\mathbf{\overline{B}} & = & \hat{\mathbf{\phi}}B+\nabla\times\left(A\hat{\mathbf{\phi}}\right)\,,\label{eq:b1}\\
\tilde{\mathbf{B}} & = & \mathbf{\nabla}\times\left(\mathbf{r}T\right)+\mathbf{\nabla}\times\mathbf{\nabla}\times\left(\mathbf{r}S\right),\label{eq:b2}
\end{eqnarray}
where $\overline{\mathbf{B}}$ and $\tilde{\mathbf{B}}$ are the axisymmetric
and nonaxisymmetric components, ${A}$, ${B}$, ${T}$ and ${S}$
are the scalar functions representing the field components, $\hat{\mathbf{\phi}}$
is the azimuthal unit vector, $\mathbf{r}$ is the radius vector,
$r$ is the radial distance, and $\theta$ is the polar angle.\textbf{
}\cite{Krause1980} showed that the only gauge transformation for
potentials T and S is a sum with the arbitrary r-dependent function.
To fix this arbitrariness they suggested the following normalization:
\begin{equation}
\int_{0}^{2\pi}\int_{-1}^{1}Sd\mu d\phi=\int_{0}^{2\pi}\int_{-1}^{1}Td\mu d\phi=0.\label{eq:norm-1}
\end{equation}

\begin{figure}
\centering \includegraphics[width=0.99\columnwidth]{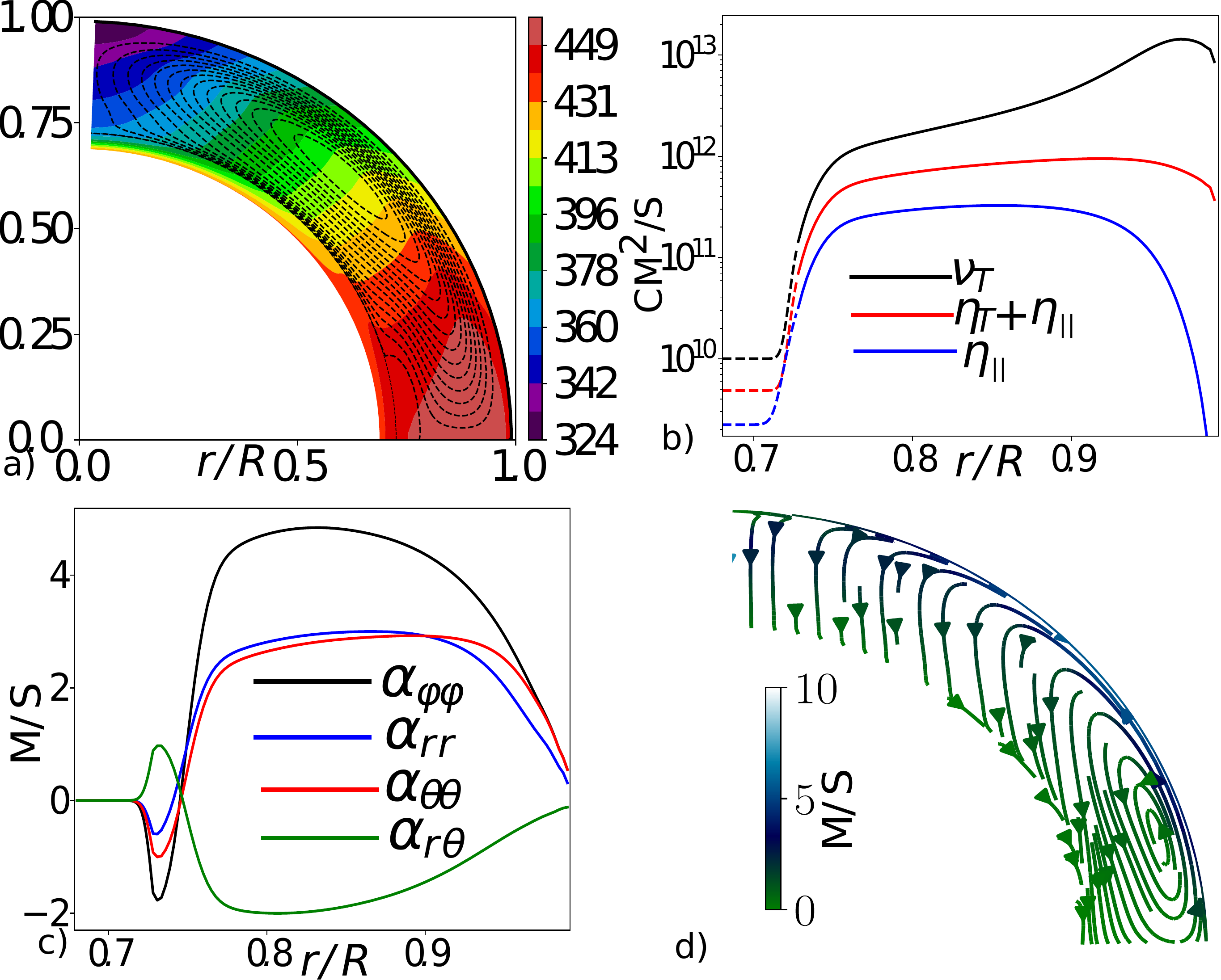} \caption{\label{fig1} a) The basic angular velocity profile and the streamlines
of the meridional circulation with the maximum circulation velocity
of 13 m/s on the surface at the latitude of 45$^{\circ}$; b) radial
profiles of the total, $\eta_{T}+\eta_{||}$, and the rotationally
induced part, $\eta_{||}$, of the eddy magnetic diffusivity and the
eddy viscosity profile; c) radial profiles of the $\alpha$-effect
tensor at the latitude of 45$^{\circ}$; and d) the effective drift
velocity due to the meridional circulation and the turbulent pumping
effect. }
\end{figure}

We are using the same formulation for the mean electromotive force
as in the recent paper by \cite{Pipin2019c}: 
\begin{eqnarray}
\mathcal{E}_{i} & = & \mathcal{E}_{i}^{(A)}+\mathcal{E}_{i}^{(P)}\,,\label{eq:EMF-1}\\
\mathcal{E}_{i}^{(A)} & = & \left(\alpha_{ij}+\gamma_{ij}\right)\left\langle B\right\rangle _{j}-\eta_{ijk}\nabla_{j}\left\langle B\right\rangle _{k},\nonumber \\
\mathcal{E}_{i}^{(P)} & = & \alpha_{\beta}\delta_{i\phi}\left\langle B\right\rangle _{\phi}-V_{\beta}\left(\hat{\boldsymbol{r}}\times\left\langle \mathbf{B}\right\rangle \right)_{i}\,,
\end{eqnarray}
where we divide the expression of the mean electromotive force into
the ``standard'' and phenomenological parts. The standard part of
$\mathcal{E}_{i}^{(A)}$ contains contributions of the $\alpha$-effect
tensor, $\alpha_{ij}$, the antisymmetric tensor\textbf{ }$\gamma_{ij}$
accounts for the turbulent pumping of the large-scale magnetic fields
due to the effects of the global rotation and density stratification,
and $\eta_{ijk}$ is the eddy magnetic diffusivity tensor. This corresponds
to the expression of the mean electromotive force employed in the
mean-field dynamo models of \cite{Pipin2019c,Pipin2020}.
The further details concerning $\mathcal{E}_{i}^{(A)}$ can be found
in the above-cited papers. Figure \ref{fig1} shows the large-scale
flow distribution, the radial profiles of the eddy magnetic diffusivity,
the $\alpha$-effect, and the effective drift velocity due to the
meridional circulation and the turbulent pumping effect.

\begin{figure}
\includegraphics[width=0.99\columnwidth]{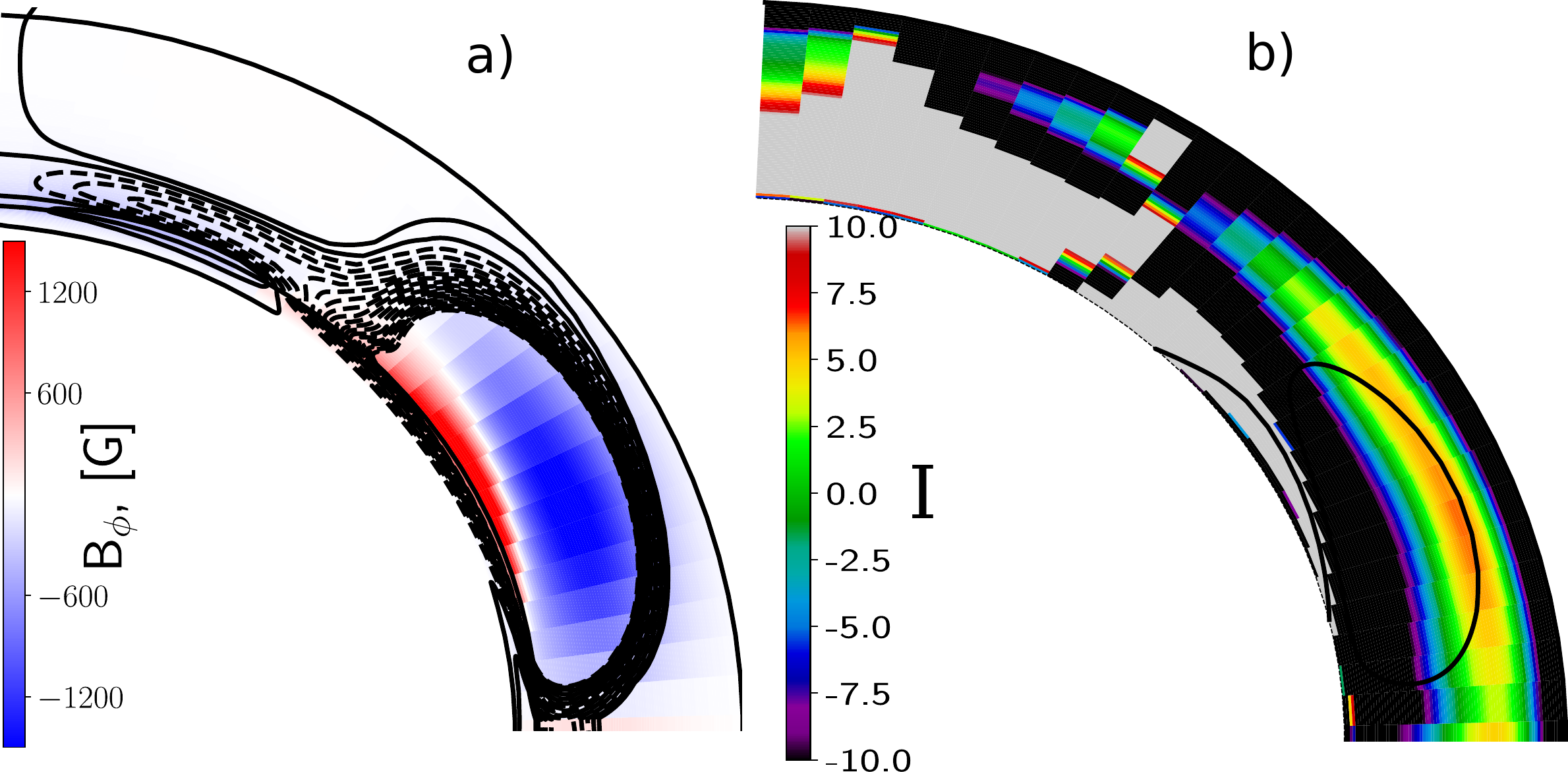} \caption{\label{fig:inst}a) Snapshot of the axisymmetric magnetic field distribution
in the convection zone in the growing phase of the magnetic cycle,
color shows the toroidal magnetic field strength and contours show
streamlines of the poloidal magnetic field (solid lines - clockwise
direction); b) color image shows the instability parameter I (see,
Eq\ref{eq:inst}), contour line shows the toroidal magnetic field
strength of 500G.}
\end{figure}

The phenomenological part of the mean-electromotive force is introduced
to account for the surface effects of the emerging magnetic active
region on the large-scale dynamo. It is assumed that the toroidal
magnetic field in the upper part of the convection zone can be buoyantly
unstable and forms magnetic bipolar regions on the solar surface.
Following \cite{Pipin2018d} we put 
\begin{eqnarray}
V_{\beta} & = & -\frac{\alpha_{MLT}u'}{\gamma}\beta^{2}K\left(\beta\right)\left(1+\xi_{\beta}(t,r,\phi,\theta)\right)\,,\label{eq:bu}
\end{eqnarray}
where $\mathrm{\alpha_{MLT}}=1.9$ is the mixing-length theory parameter,
$\gamma$ is the adiabatic law constant, $\mathbf{\boldsymbol{\Lambda}}^{(\rho)}=\boldsymbol{\nabla}\log\overline{\rho}$;
functions $f_{1,3}^{(a)}$ and $K\left(\beta\right)$ are given in
\cite{Kitchatinov1993}. The parameter $\xi_{\beta}$ determines the
magnetic buoyancy velocity perturbations. The position of the unstable
layer was computed following the consideration of \cite{Parker1979}.
It is assumed that the large-scale toroidal magnetic field becomes
unstable when its strength decreases outward faster than the mean
density does. In particular, we compute the parameter 
\begin{equation}
I=-r\frac{\partial}{\partial r}\log\left(\frac{\overline{B}}{\overline{\rho}}\right),\label{eq:inst}
\end{equation}
where $\overline{B}$ is the strength of the toroidal magnetic field
and $\overline{\rho}$ is the density profile. For the unstable part
of the toroidal magnetic field $I>0$. Note, that the magnetic buoyancy
instability can be affected by the stellar turbulence and the global
rotation \citep{Gilman1970,Acheson1978,Hughes2011,Gilman2018}. We
do not consider these effects in our simple model. Figure \ref{fig:inst}
shows snapshots of the magnetic field and the instability parameter
distributions during the growing phase of the magnetic cycle. We see,
that in our model the unstable layers of the toroidal magnetic field
are connected with the bottom of the convection zone and the subsurface
shear layer.

\begin{figure}
\includegraphics[width=0.7\columnwidth]{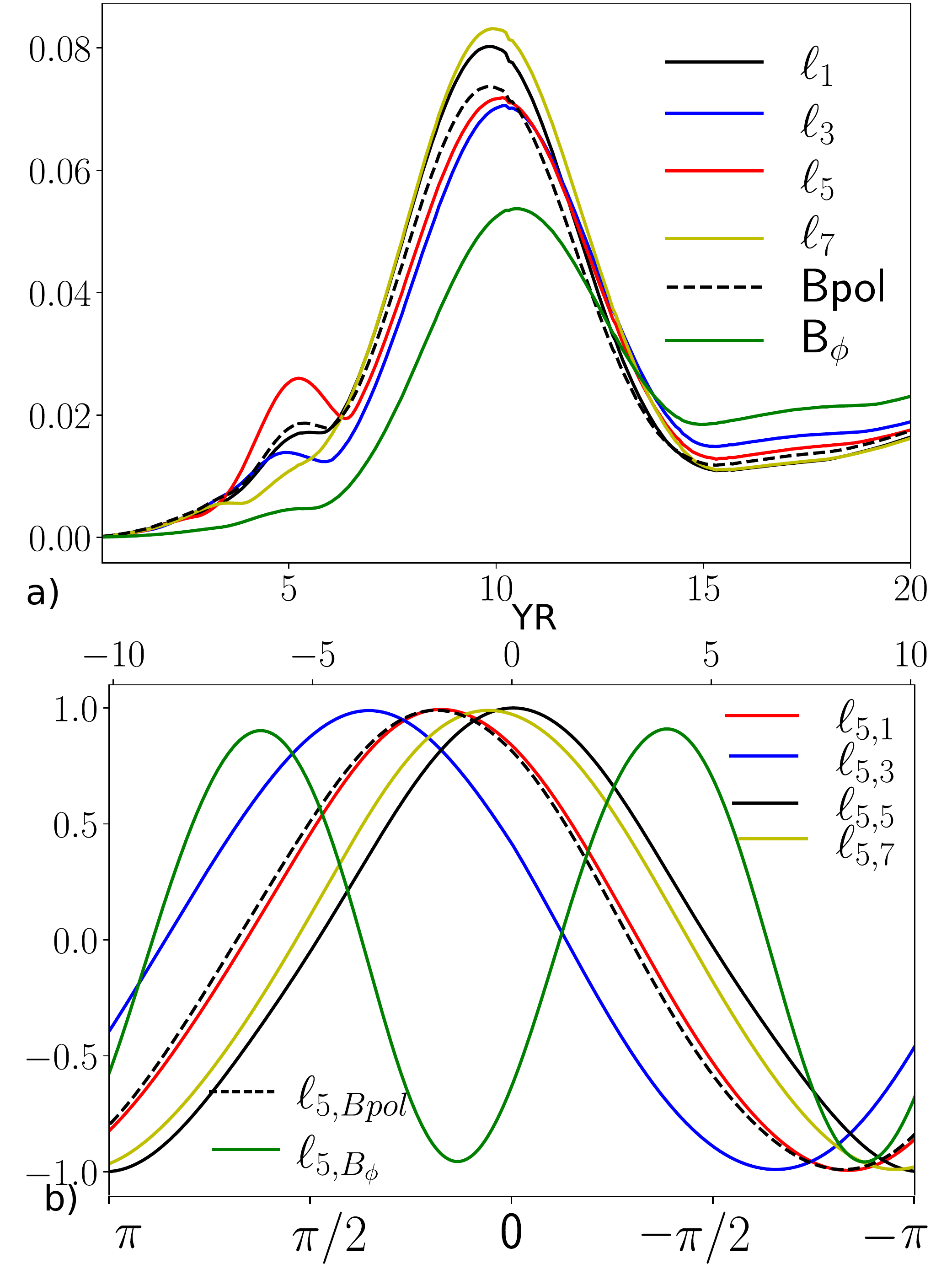} \caption{\label{fig:C2c} The same as Fig.8 for axisymmetric dynamo model.}
\end{figure}
\begin{figure*}
\includegraphics[width=0.9\textwidth]{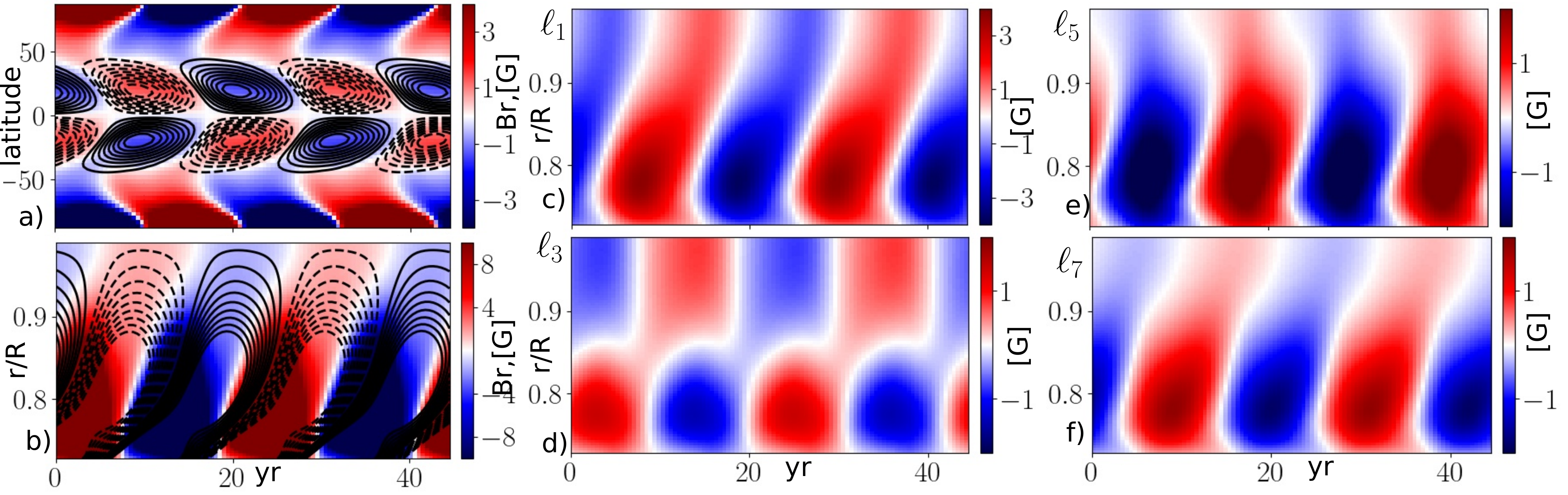} \caption{\label{fig:C2} The same as Fig.5 for axisymmetric dynamo model.}
\end{figure*}

The $\xi_{\beta}$ perturbations are initiated within the upper layer
$I>0$; initiations are random in time. They go as a Gaussian sequence
with the mean cadence equal to a month. The Gaussian sequences of
$\xi_{\beta}$ in the north and south hemispheres are independent.
Similar to \cite{Pipin2018d}, we define $\xi_{\beta}$ as follows,
\begin{equation}
\!\xi_{\beta}\!=\!C_{\beta}\!\psi\negthinspace\exp\left(\!-m_{\beta}\left(\!\sin^{2}\!\left(\!\frac{\phi-\phi_{0}}{2}\!\right)\!+\!\sin^{2}\!\left(\!\frac{\theta-\theta_{\mathrm{m}}}{2}\!\right)\!\right)\!\right)\!,\label{eq:xib}
\end{equation}
where $\psi$ is a kink-type function in radius and time, 
\begin{eqnarray}
\psi\!\! & =\!\! & \frac{1}{2}\left(\!1\!-\!\mathrm{erf}\left(50\left(r-r_{m}\right)\right)^{2}\!\right)\!\mathrm{e}\!^{{\displaystyle -\left(\frac{\tau_{0}-t}{2\tau_{0}}\right)^{2}}}\!,t\!<\!\delta t\label{eq:kink}\\
 & = & 0,t>\delta t\,,\nonumber 
\end{eqnarray}
the longitude $\phi_{0}$ is random; $\theta_{m}$ and $r_{m}$ are
the latitude and radius of the extreme points of the toroidal magnetic
field in the upper part of the convection zone, where the instability
parameter $I>0$. We restrict the active phase time, $\delta t$,
of the bipolar region evolution to one week. The emergence interval
of the bipolar region is controlled by the parameter $\tau_{0}$.
In our simulations we put it to 3 days, which corresponds to the observation
results (see, \cite{Stenflo2012a}). The parameter $C_{\beta}=200$
is chosen in such a way that the unstable part of the axisymmetric
magnetic field could emerge on the surface after time $\tau_{0}$
starting from the level r=0.9R, which corresponds to the mean level
of the magnetic buoyancy instability of the large-scale toroidal magnetic
field in the upper part of the convection zone. In this model we do
not intend to reproduce the spatial parameters of the sunspot bipolar
regions. For the sake of numerical efficiency we put $m_{\beta}=20$,
which results in large-scale bipolar regions, (see, Figure \ref{snps}).
The $\alpha$-effect will be given as follows 
\begin{equation}
\alpha_{\beta}=0.3\cos\theta V_{\beta},\label{eq:ab}
\end{equation}
where $V_{\beta}$ is determined by Eq(\ref{eq:bu}), and the coefficient
$0.3$ was chosen from numerical experiments. With our representation
of the mean electromotive force in the form of Eq(\ref{eq:EMF-1}),
the full version of the dynamo equations for the axisymmetric magnetic
field reads as follows, { 
\begin{eqnarray}
\partial_{t}A & = & \mathcal{E}_{\phi}^{(A)}+\overline{\alpha_{\beta}\left\langle B_{\phi}\right\rangle }+\frac{V_{\beta}}{r}\frac{\partial\left(rA\right)}{\partial r}+\overline{V_{\beta}\tilde{B}_{\theta}},\label{eq:At}\\
\partial_{t}B & = & \frac{\sin\theta}{r}\frac{\partial\left(r\sin\theta A,\Omega\right)}{\partial\left(r,\mu\right)}-\frac{1}{r}\frac{\partial}{\partial r}r^{2}\left(V_{\beta}B+\overline{V_{\beta}\tilde{B}_{\phi}}\right)\nonumber \\
 & + & \frac{1}{r}\frac{\partial r\mathcal{E}_{\theta}^{(A)}}{\partial r}+\frac{\sqrt{\left(1-\mu^{2}\right)}}{r}\frac{\partial\mathcal{E}_{r}^{(A)}}{\partial\mu}\,,\label{eq:Bt}
\end{eqnarray}
}where, the contribution of $\mathcal{E}^{(A)}$ stands for the standard
part of the mean-electromotive force, its details are omitted. For
the sake of brevity, the $\alpha_{\beta}$-effect and magnetic buoyancy
terms are written explicitly via the magnetic field components. Besides,
these contributions bear the effect of the nonlinear coupling between
the axisymmetric and nonaxisymmetric modes of the magnetic field.
For example, we have $\overline{\alpha\left\langle B\right\rangle _{\phi}}=\overline{\alpha}B+\overline{\tilde{\alpha}\tilde{B}_{\phi}}$
and the same is true for other similar terms. Note that the second
part of this formula, the term $\overline{\tilde{\alpha}\tilde{B}_{\phi}}$,
as well as the terms like $\overline{\alpha_{\beta}\left\langle B\right\rangle _{\phi}}$
and similar ones that are related to the magnetic buoyancy, are usually
ignored in standard mean-field dynamo models. In general, we can see
some similarity between the effect due to $\overline{\tilde{\alpha}\tilde{B}_{\phi}}$
and the non-local $\alpha$-effect employed in the flux-transport
models \citep{Cameron17}. In our approach, we explicitly take into
account the dynamics of the nonaxisymmetric magnetic field and its
averaged effect on the evolution of the large-scale magnetic field.
To get the evolution equation for the nonaxisymmetric potential $S$
we substitute Eq(\ref{eq:b2}) into Eq(\ref{eq:mfe}) and, then, calculate
the scalar product with vector $\mathbf{\hat{r}}$. Similarly, the
equation for $T$ is obtained by taking curl of Eq(\ref{eq:mfe})
and, then, the scalar product with vector $\mathbf{\hat{r}}$. The
procedure was described in detail by \cite{Krause1980}. See also
\cite{Bigazzi2004} and \cite{Pipin2015a}. The equations for the
potentials T and S are 
\begin{eqnarray}
\partial_{t}\Delta_{\Omega}S & = & \Delta_{\Omega}U^{(U)}+\Delta_{\Omega}U^{(\mathcal{E}A)}+\Delta_{\Omega}U^{(\mathcal{E}P)},\label{eq:S}\\
\partial_{t}\Delta_{\Omega}T & = & \Delta_{\Omega}V^{(U)}+\Delta_{\Omega}V^{(\mathcal{E}A)}+\Delta_{\Omega}V^{(\mathcal{EP})},\label{eq:T}
\end{eqnarray}
where we introduce new notations 
\begin{eqnarray}
\Delta_{\Omega}V^{(U)} & = & -\hat{\mathbf{r}}\cdot\boldsymbol{\nabla}\times\boldsymbol{\nabla}\times\left(\mathbf{\overline{U}}\times\mathbf{\tilde{\mathbf{B}}}\right),\label{eq:vu}\\
\Delta_{\Omega}V^{(\mathcal{E})} & = & -\hat{\mathbf{r}}\cdot\boldsymbol{\nabla}\times\boldsymbol{\nabla}\times\boldsymbol{\boldsymbol{\mathcal{E}}},\label{eq:ve}\\
\Delta_{\Omega}U^{(U)} & = & -\hat{\mathbf{r}}\cdot\boldsymbol{\nabla}\times\left(\mathbf{\overline{U}}\times\tilde{\mathbf{B}}\right),\label{eq:uu}\\
\Delta_{\Omega}U^{(\mathcal{E})} & = & -\hat{\mathbf{r}}\cdot\boldsymbol{\nabla}\times\boldsymbol{\boldsymbol{\mathcal{E}}}.\label{eq:ue}
\end{eqnarray}
We do not show details of $\Delta_{\Omega}U,V^{(U)}$ as well as the
standard parts of the $\Delta_{\Omega}U,V^{(\mathcal{E})}$ terms.
Their derivation is described in detail in the above-cited papers.
For the nonaxisymmetric parts related to the magnetic buoyancy and
$\alpha_{\beta}$ the effect reads as follows 
\begin{eqnarray}
\Delta_{\Omega}V^{(\mathcal{E}P)} & = & -\frac{1}{r\sin\theta}\frac{\partial}{\partial\phi}\frac{\partial}{\partial r}\left(r\left\langle B_{\theta}\right\rangle V_{\beta}\right)\nonumber \\
 & - & \frac{\partial}{\partial\mu}\left(\frac{\sin\theta}{r}\frac{\partial}{\partial r}\left(r\left\langle B_{\phi}\right\rangle V_{\beta}\right)\right)\\
 & - & \frac{1}{r}\frac{\partial}{\partial r}r\left[\frac{1}{\sin\theta}\frac{\partial}{\partial\phi}\alpha_{\beta}\left\langle B_{\phi}\right\rangle \right],\nonumber 
\end{eqnarray}
\begin{eqnarray*}
\Delta_{\Omega}U^{(\mathcal{E}P)} & = & -\frac{1}{\sin\theta}\frac{\partial}{\partial\phi}\left(\left\langle B_{\phi}\right\rangle V_{\beta}\right)+\frac{\partial}{\partial\mu}\left(\sin\theta\left\langle B_{\theta}\right\rangle V_{\beta}\right)\\
 & + & \frac{\partial}{\partial\mu}\alpha_{\beta}\left\langle B_{\phi}\right\rangle .
\end{eqnarray*}

\subsection*{B. Results of axisymmetric dynamo model}

{Here we reproduce some results of the run C2 from \cite{Pipin2020}.
See the above-cited paper for details of the model.}
Figure \ref{fig:C2c}a shows variations in the first four odd axisymmetric
spherical harmonics on the surface, as well as the evolution of the
mean polar magnetic field and the mean density of the toroidal field
in the subsurface layer, $r=0.9-0.99R$. The parameter $\overline{B_{\phi}}$
is considered a proxy for the sunspot number. The integral wavelets
of the low axisymmetric modes and their cross-correlations with $\ell_{5}$
mode are shown in Figure\ref{fig:C2c}b. The toroidal magnetic field
and the $\ell_{3}$ harmonic show periods of about 10.6 years. The
polar magnetic field, as well as the $\ell_{1,5,7}$ axisymmetric
harmonics show a period of about 11.3 years. The axisymmetric modes
show an auxiliary peak around 5 years. In the axisymmetric model,
the correlation of the mode $\ell_{5}$ with other axisymmetric modes,
as well as with the integral parameters $B_{pol}$ and the sunspot
proxy $B_{\phi}$ agrees qualitatively with observations (see, Figure\ref{fig:C2c}b).
The correlation between $\ell_{5}$ and $\ell_{7}$ shows the largest
difference compared to the observation results.
The time-latitude and time-radius diagrams of the large-scale magnetic
field evolution, as well as the time-radius evolution of the first
four odd axisymmetric spherical harmonics are shown in Figure \ref{fig:C2}.
The butterfly diagram of the model qualitatively agrees with observations.
At the latitude of 30$^{\circ}$, the diagram of the time-radius evolution
of the toroidal and radial magnetic field shows a drift of the magnetic
field activity toward the surface. The duration of the drift is about
half a magnetic cycle. It is interesting to note that the axisymmetric
harmonics $\ell_{1}$ and $\ell_{7}$ follow the same direction of
the drift. The $\ell_{3}$ harmonic shows a steady wave with two maxima:
one near the bottom and the other at the top of the convection zone.
The $\ell_{5}$ harmonic shows a similar tendency. However its maximum
at the top of the convection zone is less pronounced than for the
$\ell_{3}$ . In the upper part of the convection zone, $\ell_{5}$
displays an inward drift. Therefore, the model evolution of the radial
magnetic field of $\ell_{3}$ and $\ell_{5}$ harmonics can be connected
with the toroidal magnetic field in the subsurface shear layer. Naturally,
the $\ell_{5}$ maxima at the surface correspond to the maxima of
the toroidal magnetic field butterfly diagrams.

The above consideration reveals the peculiarities in the evolution
of the $\ell_{3}$ and $\ell_{5}$ axisymmetric harmonics of the radial
magnetic field. The harmonic $\ell_{3}$ displays a dynamo period
close to the period of the toroidal magnetic field parameter $\overline{B_{\phi}}$
in the subsurface layer. This harmonic is shifted ahead by about $\pi/2$
of the toroidal magnetic field evolution and has its maximum at the
surface. Therefore, the surface evolution of $\ell_{3}$ is determined
by generation of the radial magnetic field from the toroidal magnetic
field by means of the $\alpha$-effect. While the generation of $\ell_{3}$
is connected with the rising branch of the toroidal magnetic field
in the subsurface shear layer, the near-surface generation of $\ell_{5}$
is connected with the decaying $\overline{B_{\phi}}$ there. Observations
show that the periods of $\ell_{5}$ and $\overline{B_{\phi}}$ are
close. This means that there may be an additional near-surface source
of the $\alpha$-effect that produces $\ell_{5}$. We assume that
it can be due to the emerging bipolar regions. 
\end{document}